\documentclass[12pt]{article}
\usepackage[psamsfonts]{amsfonts,amssymb}
\usepackage[psamsfonts,mathscr]{eucal}
\usepackage{epsf}
\epsfverbosetrue

\newcommand{\Def}{\newcommand}
\Def{\DeF}{\renewcommand}
\Def{\Thm}{\newtheorem}

\Thm{definition}{Definition}
\Thm{lemma}{Lemma}
\Thm{proposition}{Proposition}
\Thm{theorem}{Theorem}
\Thm{remark}{Remark}

\Def{\bq}{\begin{equation}}
\Def{\eq}{\end{equation}}
\Def{\bQ}{\begin{eqnarray}}
\Def{\eQ}{\end{eqnarray}}

\DeF{\theequation}{\thesection.\arabic{equation}}

\Def{\And}{\mathrm{and}}
\Def{\ie}{\mathrm{i.e.\,}}
\Def{\implies}{\Longrightarrow}
\Def{\Log}{\mathrm{Log\,}}
\Def{\Or}{\mathrm{or}}
\Def{\where}{\mathrm{where}}
\Def{\with}{\mathrm{with}}

\Def{\CC}{\mathscr{C}}
\Def{\DD}{\mathscr{D}}
\Def{\TT}{\mathscr{T}}

\Def{\AAA}{\mathbb{A}}
\Def{\BBB}{\mathbb{B}}
\Def{\DDD}{\mathbb{D}}
\Def{\III}{\mathbb{I}}
\Def{\KKK}{\mathbb{K}}
\Def{\RRR}{\mathbb{R}}
\Def{\TTT}{\mathbb{T}}
\Def{\ZZZ}{\mathbb{Z}}

\Def{\ID}{\mathbf{1}}

\DeF{\refname}{}
\makeatletter
\DeF\tableofcontents{\@starttoc{toc}}
\makeatother

\begin{document}
\title{\vspace{-25mm}\rule{0cm}{2cm} \textbf{
Non Permanent Form Solutions \\
in the Hamiltonian Formulation \\
of Surface Water Waves}}
\author{T. Benzekri\thanks{\ \mbox{benzekri@cpt.univ-mrs.fr}}
\and
R. Lima\thanks{\ \mbox{lima@cpt.univ-mrs.fr}}
\and
M. Vittot\thanks{\ \mbox{vittot@cpt.univ-mrs.fr}}
}
\date{June 18, 1998 (revised version February 28, 2000)}

\maketitle
\thispagestyle{empty}

\begin{center}
{\Large Centre de Physique Th\'eorique \footnote{\ Unit\'e Propre de
Recherche 7061}}

\vspace{0.1cm}CNRS Luminy, case 907 -- F-13288 Marseille cedex 9 --
France

\vspace{0.5cm}
\textbf{Abstract}
\end{center}

Using the KAM method, we exhibit some solutions of a finite
dimensional approximation of the Zakharov Hamiltonian formulation of
gravity water waves (Zakharov, 1968), which are spatially periodic,
quasi-periodic in time, and not permanent form travelling waves. For
this Hamiltonian, which is the total energy of the waves, the
canonical variables are some complex quantities $ a_{n}$ and
$a_{n}^{*}$ ( $n \in \mathbb{Z}) $), which are linear combinations of the
Fourier components of the free surface elevation and the velocity
potential evaluated at the surface. We expose the method for the case
of a system with a finite number of degrees of freedom, the Zufiria
model (Zufiria, 1988), with only 3 modes interacting.

\vspace{1cm}\textbf{Keywords:} KAM theory, non-permanent waves.

\vspace{1cm}

CPT-98/P.3694

Published in the ``European Journal of Mechanics B: Fluids'',
\underline{19}, (2000) 379-390
\vspace{1.5cm}

\newpage
\vspace{5mm}

\section{Introduction}
\label{S1}
We consider the problem of waves of the free surface of an inviscid
and incompressible fluid which satisfies the Euler equation. It is
well known that this problem can be formulated as an Hamiltonian
system (Zakharov, 1968) with canonical conjugate variables \(
(\eta(x), \psi(x)) \). Here \( \eta(x) \) is the free surface
displacement above the point \( x \) of the (planar horizontal)
bottom, and \( \psi(x) \) is the velocity potential taken at the
surface \( \psi(x) = \Phi(x, \eta(x)) \) . There are infinitely many
degrees of freedom, labeled by the parameter \( x \) .

Several important numerical and analytical studies have been performed
in connection with the question of existence and stability of
travelling (or permanent form) waves. They are solutions of the form
\( \eta(x,t) = g(x - c.t) \) for some function \( g \) and constant c.
We refer in particular to the works of (Levi-Civita, 1925), (Crapper,
1957), (Longuet-Higgins, 1978), (McLean, 1980), (Chen \& Saffman,
1980), (Mackay \& Saffman, 1986), (Craig \& Sulem, 1993), (Craig \&
Worfolk, 1995), (Craig, 1996), (Bridges \& Dias, 1996) and (Debiane \&
Kharif, 1996).

From now on, we restrict ourself to the case where the variable \( x
\in \mathbb{R} \), and the solutions \( \ \eta, \psi\ \) are required to
satisfy periodic boundary conditions, in \( x \) (in \( [0, 2\pi] \)
for instance).

Of particular interest for searching the solutions of an Hamiltonian
system, the Birkhoff reduction to normal forms (Birkhoff, 1927) is
known to be a powerful method of analysis. It even gives rise, for
some cases of the water waves problem, to integrable reduced
Hamiltonians as it was shown in (Dyachenko \& Zakharov, 1994). However
in (Craig \& Worfolk, 1995), the authors shown that contrary to the
conjecture stated in the preceding paper, the fifth order Birkhoff
form of the full Zakharov Hamiltonian is no more integrable due to the
presence of resonant terms. It is well known that such terms will
prevent the Birkhoff iterative process to converge (problem of small
divisors): see for instance (Siegel \& Moser, 1971).

On the other hand, the KAM theory, (Kolmogorov, 1954), (Arnold, 1963)
and (Moser, 1963) indicates that under some conditions, an Hamiltonian
system treated as a quasi-integrable one exhibits a great number of
quasi-periodic solutions on invariant tori, provided that the
integrable part of the Hamiltonian system is non degenerate, and that
the perturbation is small enough. By quasi-integrable we mean that the
Hamiltonian can be written as an integrable one, added with a small
quantities in a sense to be precised. We also need that some
parameters (or initial conditions) verify a non-resonance condition of
Diophantine type: i.e. they must be in some Cantor set.

In this paper we use the KAM method to prove the existence of a
solution of a simplified, non-integrable model of surface gravity
waves in deep water, introduced by (Zufiria, 1987). As usual in KAM
theory, we build a canonical transformation that takes the initial
Hamiltonian into an integrable one, at least locally in the phase
space. In this way it is possible to write a trajectory of the initial
Hamiltonian system for each initial non-resonant condition. It is easy
to show, by inspection, that such trajectories are not travelling
waves. It will be clear that the method can be used for any other
truncated (finite dimensional) form of the Zakharov Hamiltonian.

\vspace{0.5cm}This paper is organized as follows:

In section 2, we describe the Zakharov Hamiltonian for the gravity
water waves and we write a truncated approximation of it in
action-angle coordinates. As mentionned above, we exemplify the
technics on the Zufiria 3-modes interaction model.

Section 3 is devoted to the description of the KAM strategy in the
general case. We identify a part of the Hamiltonian which is small and
thus which can be removed by a canonical transformation. And the
remaining part of the Hamiltonian has a trivial trajectory. This
method builds the unknown transformation by an implicit equation. This
change of variables is applied to the set of non-resonant Diophantine
initial conditions in the neighborhood of a point in the phase space.
Then we obtain a trajectory of the original perturbed Hamiltonian, in
the initial physical coordinates, by computing the inverse of this
change of variables.

In section 4 we apply the KAM method to the Zufiria model: we exhibit
a part of the Hamiltonian which is small for a non-empty set of
initial conditions. We also verify that the obtained trajectories are
not travelling waves (or permanent form) solutions.

Section 5 is a brief conclusion along with some open questions and
work in progress.

Let us mention that an alternative starting hamiltonian would be any
finite-order truncation of the ones given in (Dyachenko, 1995),
(Craig, 1996) for deep water waves or (Craig, 1993) for the finite
depth case. We leave this task to further work since it would require
much more algebra, even though the final result would be physically
more accurate.

\section{The Zakharov \& Zufiria Hamiltonians}
\label{S2}
In this section we shall remind some well-known facts of the
Hamiltonian formalism of surface water waves (Zakharov, 1968). We will
study a truncated version of it.

Let \( \eta(x) \) be the elevation of the surface wave, \( \Phi(x,y)
\) the velocity potential and \( \psi(x) \) be the velocity potential
evaluated at the surface. The dynamical variables are \( (\eta , \psi)
\), canonically conjugated. We only consider the case where the bottom
is at an infinite depth.

Then the Hamiltonian \( H(\eta, \psi) \) is the total energy of the
system given as the sum of the kinetic energy \( H_{e} \) and the
potential energy \( H_{p} \), where:
\begin{equation}
\label{001}
H_{e} = \frac{1}{2} \int_{0}^{2\pi} dx \int_{-\infty}^{\eta(x)} dy\
\biggl[ \Bigl( \partial_{x} \Phi(x,y) \Bigr)^{2} + \Bigl( \partial_{y}
\Phi(x,y) \Bigr)^{2} \biggr]
\end{equation}
and
\begin{equation}
\label{002}
H_{p} = \frac{1}{2} \int_{0}^{2\pi} dx\ {\eta}^{2}(x)
\end{equation}
Let us first introduce the Fourier representation of \( \eta(x)\ \And\
\psi(x) \):
\begin{eqnarray}
\label{003}
&& \eta(x) = \sum_{k \in \mathbb{Z}} \hat{\eta}_{k} e^{ikx}\ \ \mbox{with} \ \
\hat{\eta}_{k} = \hat{\eta}_{-k}^{*} \\
&& \psi(x) = \sum_{k \in \mathbb{Z}} \hat{\psi}_{k} e^{ikx}\ \ \mbox{with} \ \
\hat{\psi}_{k} = \hat{\psi}_{-k}^{*}
\end{eqnarray}
along with the new variables:
\begin{eqnarray}
\label{004}
&& \hat{a}_{k} = (\hat{\eta}_{k} |k|^{-\frac{1}{4}} + i \hat{\psi}_{k}
|k|^{\frac{1} {4}})/\sqrt{2} \\
&& \mbox{ie} \cr
&& \hat{\eta}_{k} = (\hat{a}_{k} + \hat{a}_{-k}^{*})
{|k|}^{\frac{1}{4}}/\sqrt{2} \label{005} \\
&& \hat{\psi}_{k} = (\hat{a}_{k} - \hat{a}_{-k}^{*})
{|k|}^{-\frac{1}{4}} /(i\sqrt{2})
\label{039}
\end{eqnarray}
The dynamical variables are now \( (\hat{\eta} , \hat{\psi}) \), still
canonically conjugated, or also \( (\hat{a}_{k}, \hat{a}_{k}^{*}) \).
The Hamiltonian is then reduced to a system of simple equations:
\begin{equation}
\label{006}
\partial_{t}{\hat{a}}_{k} = - i \frac{\partial
H}{\partial{\hat{a}}_{k}^{*}} \ \ k \in \mathbb{Z}
\end{equation}
By definition, \( \eta(x,t) \) is a travelling wave if and only if
there exists a function \( g \) and a constant c such that \(
\eta(x,t) = g(x -c.t) \). This is equivalent to:
\begin{equation}
\label{041}
\partial_{t} \eta = -c.\partial_{x} \eta\ \ \mbox{or}\ \ \partial_{t}
\hat{\eta}_{k} = -i.k.c.\hat{\eta}_{k}
\end{equation}
or to the requirement that the function \( \hat{\eta}_{k}(t) \)
verifies:
\begin{equation}
\label{052}
\forall k,t \ \ \ 
\ln(\hat{\eta}_{k}(t)) = e_{k} -i.k.c.t,
\end{equation}
for some constants \( e_{k} \).

The expansion of the Hamiltonian in power series of the variables \(
\hat{a} \) is formally given by (Krasitskii, 1994).

Many models which are derived from the Zakharov Hamiltonian system
have been used to study the time evolution of spatially periodic
waves, see (Stiasnie \& Shemer, 1984), (Zufiria, 1987) and (Badulin,
Shrira, Ioualalen \& Kharif, 1995). In this paper, we choose the 3
waves interaction model introduced by Zufiria, where he assumes that
the only non-zero variables are \( \hat{a}_{1}, \hat{a}_{2},
\hat{a}_{3} \) the others being frozen to 0. Due to the existence of
an extra integral of motion (the total momentum (\ref{008}), besides
the Hamiltonian itself), this is known to be the minimal model that
may exhibit a chaotic behavior. Zufiria gives an expansion of the
Hamiltonian up to the 4th order of waves amplitudes \( \hat{a} \). We
can rewrite it as follows:
\begin{eqnarray}
\label{007}
&& H = \hat{a}_{1} \hat{a}_{1}^{*} + 2^\frac{1}{2} \hat{a}_{2}
\hat{a}_{2}^{*} + 3^\frac{1}{2} \hat{a}_{3} \hat{a}_{3}^{*}
+2^{\frac{1}{2}} \gamma_{6} \biggl( \hat{a}_{1} \hat{a}_{2}
\hat{a}_{3}^{*} + \hat{a}_{1}^{*} \hat{a}_{2}^{*} \hat{a}_{3} \biggr)
+ 2^{\frac{1}{2}} \gamma_{7} \Bigl( \hat{a}_{1}^{2} \hat{a}_{2}^{*} +
\hat{a}_{1}^{*2} \hat{a}_{2} \Bigr)\ + \cr
&& \hat{a}_{1}
\hat{a}_{1}^{*} \biggl( \frac{1}{8} \hat{a}_{1} \hat{a}_{1}^{*} + 2
w_{1} \hat{a}_{2} \hat{a}_{2}^{*} + w_{2} \hat{a}_{3} \hat{a}_{3}^{*}
\biggr) +\ \hat{a}_{2} \hat{a}_{2}^{*} \Bigl( \hat{a}_{2}
\hat{a}_{2}^{*} + 2 w_{3} \hat{a}_{3} \hat{a}_{3}^{*} \Bigr) +
\frac{27}{8} \Bigl( \hat{a}_{3} \hat{a}_{3}^{*} \Bigr)^{2}\ + \cr
&& \ 2\gamma_{8} \Bigl( \hat{a}_{1} \hat{a}_{2}^{*2} \hat{a}_{3} +
\hat{a}_{1}^{*} {\hat{a}_{2}}^{2} \hat{a}_{3}^{*} \Bigr) - \gamma_{9}
\biggl( \hat{a}_{1}^{3} \hat{a}_{3}^{*} + \hat{a}_{1}^{*3} \hat{a}_{3}
\biggr) \\
&& \mbox{where}\ \ \ \gamma_{6} = \Bigl(\frac{3}{8}\Bigr)^{\frac{1}{4}}\ ,\ \
\gamma_{7} = 2^{-{\frac{9}{4}}}\ ,\ \ w_{1} = {\frac{1}{2}} + {\frac{\sqrt{2}}{8}  
}\ ,\ \ w_{2} = {{3 + \sqrt{3}} \over 2} \\
&& w_{3} = 3 + \sqrt{\frac{3}{8}}\ ,\ \ \gamma_{8} = 2^{-{\frac{7}{2}}}.3^{-{\frac{1}{4}}}.(2 + \sqrt{3} + \sqrt{8})\ ,\ \ \gamma_{9} =
{3^{\frac{1}{4}} \over 8}
\label{009}
\end{eqnarray}
Zufiria also notices that this Hamiltonian retains all the resonant
terms up to order 4, and therefore it can be viewed as the normal form
in 3 degrees of freedom.

For the water waves equation, beside the Hamiltonian, there is an other
conserved quantity which is the total horizontal momentum given by:
\begin{equation}
\label{008}
\mathbb{I} = \sum_{k = 1}^{3}k\ \hat{a}_{k} \hat{a}_{k}^{*}
\end{equation}
(obviously the case \( \mathbb{I} = 0 \) corresponds to a flat surface, for
this Zufiria model, where the sum is only on positives k). Indeed
(\ref{008}) is a conserved quantity as it can be checked by derivation
and the help of the hamiltonian equation (\ref{006}). We can also
rederive this fact as follows. Let us introduce the action-angle
variables with the following canonical transformation from \(
(\hat{a}_{1}, \hat{a}_{1}^{*}, \hat{a}_{2}, \hat{a}_{2}^{*},
\hat{a}_{3}, \hat{a}_{3}^{*}) \) to \( (A, \theta, \mathbb{I}, \xi, B,
\varphi) \) defined by:
\begin{eqnarray}
\label{013}
&& \hat{a}_{1} = \sqrt{{a} + A}\ e^{i(\theta + \xi)} \\
&& \hat{a}_{2} = {1 \over \sqrt{2}} \sqrt{\mathbb{I} - a - 3b - A - 3B}\
e^{2i\xi} \\
&& \hat{a}_{3} = \sqrt{{b} + B}\ e^{i(\varphi + 3\xi)} \label{010}
\label{045} \\
&& \mbox{with} \ \ {a} \geq 0, \ \ {b} \geq 0,\ \ {a} + 3 {b} \leq \mathbb{I} \\
&& \And\ A \geq -{a},\ B \geq -{b},\ A + 3 B \leq \mathbb{I} - {a} - 3 {b}
\end{eqnarray}
The cases $ (a = \mathbb{I}, b = 0)$ or $(a = b = 0)$ or $(a = 0, b =
\mathbb{I}/3)$ correspond to the travelling waves of class 1, 2 or 3
(respectively) of the Zufiria classification. By studying the
stability of these solutions, he finds that the most important
solution, from a dynamical point of view, is the one of class 2.
Indeed, for a certain values of the parameter \( \mathbb{I} \), the
stability of this solution changes.

The change of variables (\ref{013}-\ref{045}) leads to a new
Hamiltonian, still denoted by H:

\[
H = c + \omega_{1} A + \omega_{3} B - \gamma_{3} A B - \gamma_{4}
A^{2} - \gamma_{5} B^{2}\ +
\]
\[
2 \Bigl[ ({a} + A) (\mathbb{I} - {a} - 3 {b} - A - 3 B) \Bigr]^{1 \over 2}
\biggl[ ({b} + B)^{1 \over 2} \gamma_{6} \cos(\theta - \varphi) + ({a}
+ A)^{1 \over 2} \gamma_{7} \cos(2\theta) \biggr]\ +
\]
\begin{equation}
\label{014}
2 \Bigl[ ({a} + A) ({b} + B) \Bigr]^{1 \over 2} \biggl[ (\mathbb{I} - {a} -
3 {b} - A - 3 B) \gamma_{8} \cos(\theta + \varphi) - ({a} + A)
\gamma_{9} \cos(3\theta - \varphi) \biggr]
\end{equation}
where:
\begin{eqnarray}
\label{047}
&& c = \gamma_{1} + \mathbb{I} \gamma_{2} + {\mathbb{I}^{2} \over 4} \\
&& \gamma_{1} = a \tilde{\omega}_{1} - b \tilde{\omega}_{3} + a b
\gamma_{3} + a^{2} \gamma_{4} + b^{2} \gamma_{5} \\
&& \gamma_{2} = {1 \over \sqrt{2}} + a \hat{\omega}_{1} +
\hat{\omega}_{3} b \\
&& \gamma_{3} = {\sqrt{3} \over 2} \Bigl( \sqrt{3} - 1 + {1 \over
\sqrt{2}} + \sqrt{3 \over 8} \Bigr) \\
&& \gamma_{4} = {\sqrt{2} + 1 \over 8} \\
&& \gamma_{5} = 3 \Bigl( {9 \over 8} + \sqrt{3 \over 8} \Bigr) \\
&& \omega_{1} = 2 a \gamma_{4} + b \gamma_{3} + \hat{\omega}_{1} \mathbb{I}
+ \tilde{\omega}_{1} \label{061} \\
&& \omega_{3} = a \gamma_{3} + 2 b \gamma_{5} + \hat{\omega}_{3} \mathbb{I}
- \tilde{\omega}_{3} \label{062} \\
&& \hat{\omega}_{1} = {1 \over 4\sqrt{2}}\quad , \quad
\hat{\omega}_{3} = {3 \over 2} \sqrt{3 \over 8} \\
&& \tilde{\omega}_{1} = 1 - {1 \over \sqrt{2}}\quad , \quad
\tilde{\omega}_{3} = \sqrt{3} + {3 \over \sqrt{2}}
\label{048}
\end{eqnarray}
From (\ref{014}) the Hessian of the hamiltonian is non-degenerate in
action variables. This will be used in section 4 in order to apply KAM
method.

Let us notice that \( H \) is independent of \( \xi \) and hence we
see again that \( \mathbb{I} \) is a constant of motion.

\section{The KAM method}
\label{S3}
\setcounter{equation}{0}

We want a canonical transformation which turns an Hamiltonian written
in action-angle variables, into another one for which we can compute
an exact trajectory. And so by inversing this transformation we will
get an exact trajectory of the initial Hamiltonian system.

So let us consider an Hamiltonian \( H \) with L degrees of freedom in
action-angle variables:
\begin{equation}
\label{015}
(A,\theta) \in \mathbb{A} \times \mathbb{T}^{L}
\end{equation}
where \( \mathbb{T} \) is the torus \( \mathbb{R}/2\pi \mathbb{Z} \) and the domain \(
\mathbb{A} \) is a neighbourhood of a point \( A_{0} \in \mathbb{R}^{L}.\ A_{0} \)
can be taken to be 0 by translation in the action variable.

Assume that \( H \) is of class \( \mathscr{C}^{2} \) with respect to A, and
let us write \( H(A)(\theta) \) instead of \( H(A,\theta) \): this
means that \( H(A) \) is a function of \( \theta \). The Taylor
expansion of \( H \) around \( A = 0 \) is then:
\begin{equation}
\label{017}
{H}(A)(\theta) = c + \omega. A - f(\theta) - V(\theta). A - q(A)
(\theta)
\end{equation}
where \( c \) is some scalar constant, \( \omega \) is some vectorial
constant \( \in \mathbb{R}^{L} \). Here \( f, V \) and \( q(A) \) are
functions of \( \theta \) with values respectively in \( \mathbb{R} ,\
\mathbb{R}^{L},\ \mathbb{R} \).

We denote by \( ' \) the derivative with respect to \( A \), and by \(
\partial = \partial_{\theta} \). So that we have \( q(0) = q'(0) = 0
\) (q is of order 2 in A). We assume that \( \mathscr{D}^{0}[q''(0)] \) is an
invertible matrix from \( \mathbb{R}^{L} \) into itself, where \( \mathscr{D}^{0}(f)
\) is the average of a function \( f \) over the angles, and:
\begin{equation}
\label{026}
\mathscr{D}^{*} \equiv \mathbf{1} - \mathscr{D}^{0}
\end{equation}
with \( \mathbf{1} \) being the identity matrix.

If \( f = V = 0 \), the Hamiltonian equations of motions are:
\begin{eqnarray}
&& \partial_{t} A = -\partial H = -\partial q(A) \\
&& \partial_{t} \theta = H' = \omega + q'(A)
\end{eqnarray}
When A is taken to be zero, we get a trivial equation. And the flow
is:
\begin{equation}
\left(
\begin{array}{c} 0 \\ \theta \end{array} \right) \to \left(
\begin{array}{c} 0 \\ \theta + \omega t \end{array} \right)
\end{equation}
for any initial condition \( \theta \).

This was the ``integrable case'', where \( f = V = 0 \). Now let us
consider the ``quasi-integrable case'', i.e. we assume that \( f \)
and \( V \) are small quantities and we define: \( \varepsilon =
\max(|f| , |V|) \) for some norm. So the hamiltonian we investigate is
quasi-integrable near \( A=0 \). To avoid ambiguities we make the
hypothesis:
\begin{equation}
\label{049}
\mathscr{D}^{0}(f) = \mathscr{D}^{0}(V) = 0
\end{equation}
Indeed the quantities \( \mathscr{D}^{0}(f)\ \And\ \mathscr{D}^{0}(V) \) can be added
to \( c\ \And\ \omega \) respectively.

We want a diffeomorphism T from \( \mathbb{A} \times \mathbb{T}^{L} \) to some
domain in \( \mathbb{R}^{L} \times \mathbb{T}^{L} \) which transforms the
Hamiltonian \( H \) in another one for which we can exhibit some
explicit trajectories.

Let us define the translation operator \( \mathscr{T} \) acting on the action
variable by:
\begin{equation}
\label{021}
\forall A \in \mathbb{R}^{L} \ \ \forall Z \in \mathbb{R}^{L} \ \ [\mathscr{T}(Z) H](A) =
H(A + Z)
\end{equation}
still valid when the members of this equality are functions of \(
\theta \). Let us also define the dilation operator \( \mathbb{D} \) (acting
on the action variable) by:
\begin{equation}
\label{022}
\forall A \in \mathbb{R}^{L} \ \ \forall {\mathscr{C}} \in L(\mathbb{R}^{L},\mathbb{R}^{L}) \ \
[\mathbb{D}(\mathscr{C}) H](A) = H({\mathscr{C}}A)
\end{equation}
so that \( \mathbb{D}(\mathscr{C})^{-1} = \mathbb{D}(\mathscr{C}^{-1}) \).

Let us choose the following ansatz for \( T \):
\begin{eqnarray}
\label{019}
&& T = T_{2} . T_{1}\ \ \mbox{with} \\
&& T_{1} = \mathscr{T} (\dot{\alpha} - Z) \\
&& T_{2} = (1 + G \partial)^{-1} \mathbb{D} (\mathbf{1} + \dot{G} )
\end{eqnarray}
and where:
\begin{equation}
\label{020}
Z \in \mathbb{R}^{L},\ \ \alpha: \mathbb{T}^{L} \longrightarrow \mathbb{R},\ \ G:
\mathbb{T}^{L} \longrightarrow \mathbb{R}^{L},\ \ \dot{\alpha} \equiv \partial
\alpha,\ \ \dot{G} \equiv \partial G
\end{equation}
so that Z is a constant vector. Let us emphasize that we denote by a
dot the derivative with respect to \( \theta \). Of course we assume
that \( \dot{G} \) is small enough to have \( \mathbf{1} + \dot{G} \)
invertible.

Notice that \( T_{1}, T_{2} \) (and so \( T \) ) are canonical
transformations. Indeed:
\begin{eqnarray}
&& T_{1}A = A + \dot{\alpha} - Z\ \implies\ d(T_{1}A) = dA +
\ddot{\alpha} d\theta \\
&& T_{1} \theta = \theta\ \implies\ d(T_{1}\theta) \wedge d(T_{1}A) =
d\theta \wedge (dA + \ddot{\alpha} d\theta) = d\theta \wedge dA
\label{040}
\end{eqnarray}
since \( d\theta \wedge d\theta = 0 \). And instead to prove that \(
T_{2} \) is canonical, let us rather prove this for \( T_{3} \equiv
T_{2}^{-1} = \mathbb{D} (\mathbf{1} + \dot{G})^{-1} (1 + G \partial) \):
\begin{eqnarray}
&& T_{3}A = (\mathbf{1} + \dot{G})^{-1}A\ \implies\ d(T_{3}A) = (\mathbf{1} +
\dot{G})^{-1} dA + (\ldots) d\theta \\
&& T_{3} \theta = (1 + G \partial)\theta = \theta + G\ \implies\
d(T_{3}\theta) = d\theta (\mathbf{1} + \dot{G}) \label{018} \\
&& d(T_{3}\theta) \wedge d(T_{3}A) = d\theta (\mathbf{1} + \dot{G}) \wedge
(\mathbf{1} + \dot{G} )^{-1} dA +d\theta (\mathbf{1} + \dot{G}) \wedge (\ldots)
d\theta = d\theta \wedge dA \nonumber
\end{eqnarray}
Q.E.D.

Notice also that:
\begin{eqnarray}
\label{056}
T^{-1} \left( \begin{array}{c} A \\ \theta \end{array} \right) =
\left( \begin{array}{c} (\mathbf{1} + \dot{G})^{-1} (A + Z - \dot{\alpha}) \\
\theta + G \end{array} \right)
\end{eqnarray}
The transformed Hamiltonian \( \hat{H} = TH \):
\begin{equation}
\label{030}
\hat{H}(A) = (1 + G \partial)^{-1}H[(\mathbf{1} + \dot{G}) A + \dot{\alpha} -
Z]
\end{equation}
can be written similarly as (\ref{017}):
\begin{eqnarray}
\label{023}
&& \hat{H}(A) = \hat{c} + \hat{\omega}.A - \hat{f} - \hat{V}.A -
\hat{q}(A) \\
&& \mbox{where}\ \ \ \hat{c} = c - \omega.Z - \mathscr{D}^{0} \Bigl[ V.\dot{\alpha}
+ q(\dot{\alpha} - Z) \Bigr] \\
&& \hat{\omega} = \omega - \mathscr{D}^{0} \Bigl[ V.\dot{G} + q'(\dot{\alpha}
- Z).(\mathbf{1} + \dot{G}) \Bigr] \\
&& \hat{f} = (1 + G \partial)^{-1} \mathscr{D}^{*} \Bigl[ f -
\omega.\dot{\alpha} + V.\dot{\alpha} - V.Z + q(\dot{\alpha} - Z)
\Bigr] \\
&& \hat{V} = (1 + G \partial)^{-1} \mathscr{D}^{*} \Bigl[ V.(\mathbf{1} + \dot{G}) -
\omega.\dot{G} + q'(\dot{\alpha} - Z).(\mathbf{1} + \dot{G}) \Bigr]
\end{eqnarray}
and
\begin{equation}
\label{024}
\hat{q}(A) = (1 + G \partial)^{-1} \mathscr{D}^{*} \biggl[ q \Bigl(
\dot{\alpha} - Z + (\mathbf{1} + \dot{G}).A \Bigr) - q(\dot{\alpha} - Z) -
q'(\dot{\alpha} - Z).(\mathbf{1} + \dot{G}).A \biggr]
\end{equation}
Obviously:
\begin{equation}
\label{025}
\hat{q}(0) = \hat{q}'(0) = 0
\end{equation}
Hence if we manage to find \( \alpha,\ G,\ Z \) such that \( \hat{f} =
\hat{V} = 0 \) then we will get a trajectory of the transformed
hamiltonian \( \hat{H} \) since the initial data \( A = 0 \) remains
forever frozen to 0 under the evolution of the special hamiltonian:
\begin{equation}
\label{028}
\hat{H}(A)(\theta) = \hat{c} + \hat{\omega}.A - \hat{q}(A)(\theta)
\end{equation}
for which the flow is:
\begin{equation}
\left( \begin{array}{c} 0 \\ \theta
\end{array} \right) \to \left( \begin{array}{c} 0 \\ \theta +
\hat{\omega}t \end{array} \right)
\label{029}
\end{equation}
for any initial condition \( \theta \).

Using the inverse transformation \( T^{-1} \), we get a trajectory of
the initial Hamiltonian \( H \):
\begin{eqnarray}
\label{031}
T^{-1} \left( \begin{array}{c} 0 \\ \theta \end{array} \right) \to
T^{-1} \left( \begin{array}{c} 0 \\ \theta + \hat{\omega}t \end{array}
\right)
\end{eqnarray}
where:
\begin{eqnarray}
\label{055}
T^{-1} \left( \begin{array}{c} 0 \\ \theta \end{array} \right) =
\left( \begin{array}{c} (\mathbf{1} + \dot{G})^{-1} (Z - \dot{\alpha}) \\
\theta + G
\end{array}
\right)
\end{eqnarray}
In order to have \( \hat{f} = \hat{V} = 0 \) we require \( \alpha,\ G\
\&\ Z \) to verify:
\begin{eqnarray}
\label{027}
&& \omega.\dot{\alpha} = \mathscr{D}^{*} \biggl[ f + V.\dot{\alpha} - V.Z +
q(\dot{\alpha} - Z) \biggr] \\
&& \omega.\dot{G} = \mathscr{D}^{*} \biggl[ V.(1 + \dot{G}) + q'(\dot{\alpha}
- Z).(\mathbf{1} + \dot{G}) \biggr] \label{054}
\end{eqnarray}
Let us define the formal inverse of the operator \( \omega. \partial
\) by:
\begin{equation}
\label{034}
\Gamma \equiv (\omega. \partial)^{-1}. \mathscr{D}^{*}
\end{equation}
which is defined on any function of the angles, by its action on the
exponential function:
\begin{eqnarray}
\label{035}
\forall \nu \in \mathbb{Z}^{L} \setminus \{0\}\ \ \ \ \Gamma. e_{\nu} \equiv
(i. \omega. \nu)^{-1}. e_{\nu}\ \ \ \ \mbox{where}\ \ \ \ e_{\nu}(\theta) =
e^{i \nu.\theta}
\end{eqnarray}
and \( \Gamma.1=0 \).

Of course we need to assume an arithmetic condition on \( \omega \)
namely that it satisfies a Diophantine condition:
\begin{equation}
\label{016}
\forall \nu \in \mathbb{Z}^{L} \setminus \{0\}\ \ \ \ |\omega.\nu| \geq
\gamma |\nu|^{-\tau}
\end{equation}
for some numbers \( \gamma > 0,\ \tau > L - 1 \). The set of such
frequencies has a positive Lebesgue measure (if \( \gamma \) is small
enough), as is well known.

Then (\ref{027}), (\ref{054}) can be rewritten formally as:
\begin{eqnarray}
\label{057}
&& \alpha = \Gamma \biggl[ f + V.\dot{\alpha} - V.Z + q(\dot{\alpha} -
Z) \biggr] \\
&& G = \Gamma \biggl[ V.(\mathbf{1} + \dot{G}) + q'(\dot{\alpha} - Z).(\mathbf{1} +
\dot{G}) \biggr]\ \ \mbox{ie:} \label{032} \\
&& G = {1 \over {1 - \Gamma [V + q'(\dot{\alpha} - Z)] \partial}}\ \
\Gamma [V + q'(\dot{\alpha} - Z)]
\label{033}
\end{eqnarray}
Notice that \( \alpha \) is defined by a fixed-point equation
(\ref{057}) \( \alpha = \varepsilon Y(\alpha) \). We will adjust the
(vector) parameter \( Z \) by requiring that \( \hat{\omega} = \omega
\), i.e.:
\begin{equation}
\mathscr{D}^{0}\Bigl[V.\dot{G} + q'(\dot{\alpha} - Z).(\mathbf{1} + \dot{G})\Bigr]
= 0
\label{038}
\end{equation}
At order \( \varepsilon \), the equation for Z is:
\begin{equation}
\label{011}
\mathscr{D}^{0} \Bigl[ q''(0)(\dot{\alpha} - Z) \Bigr] = 0
\end{equation}
and so:
\begin{equation}
\label{012}
Z= \Bigl[ \mathscr{D}^{0}[q''(0)] \Bigr]^{-1} \mathscr{D}^{0}[q''(0) \Gamma \dot{f}] +
O({\varepsilon}^{2})
\end{equation}
To solve (\ref{057}) \& (\ref{038}), we can use a particular version
of KAM method (Vittot, 1994). See also (Gallavotti, 1994).

\section{The Zufiria model as a quasi-integrable Hamiltonian}
\label{S4}
\setcounter{equation}{0}

Here we apply the previous KAM method to the Zufiria Hamiltonian
(\ref{014}). We consider \( \mathbb{I} \) as a parameter (since \( \mathbb{I} \)
is constant on any trajectory). So H becomes an Hamiltonian in 2
degrees of freedoms \( (A,\theta,B,\varphi) \), ie \( L = 2 \). We
exclude the cases $\mathbb{I} = 0,$ or $a = 0,$ or $ b = 0,$ or $ a + 3b
= \mathbb{I}$. So we can write \( H \) under the form (\ref{017}) with:
\begin{eqnarray}
\label{042}
&& f(\theta,\varphi) = 2\Bigl[ {a} (\mathbb{I} - {a} - 3 {b}) \Bigr]^{1
\over 2} \biggl[ {b}^{1 \over 2} \gamma_{6} \cos(\theta - \varphi) +
2{a}^{1 \over 2} \gamma_{7} \cos(2\theta) \biggr]\ + \cr
&& ({a} {b})^{1 \over 2} \biggl[ (\mathbb{I} - {a} - 3 {b}) \gamma_{8}
\cos(\theta + \varphi) - {a} \gamma_{9} \cos(3\theta - \varphi)
\biggr]
\end{eqnarray}
Likewise \( V = V = (V_{1}, V_{3}) \) \ with:
\begin{eqnarray}
\label{043}
&& V_{1}(\theta,\varphi) = [ {b} / {a} ]^{1 \over 2} { (\mathbb{I} - 2{a} -
3{b}) \over \bigl[ {\mathbb{I} - {a} - 3{b}} \bigr]^{1 \over 2}} \gamma_{6}
\cos(\theta - \varphi)\ + [ {b} / {a} ]^{1 \over 2} (\mathbb{I} - 3{a} -
3{b}) \gamma_{8} \cos(\theta + \varphi)\ - \cr
&& 3 ({a} {b})^{1 \over 2} \gamma_{9} \cos(3\theta - \varphi)\ +\ {
(2\mathbb{I} - 3{a} - 6{b}) \over [{\mathbb{I} - {a} - 3{b}} ]^{1 \over 2}}
\gamma_{7} \cos(2\theta)
\end{eqnarray}
and:
\begin{eqnarray}
\label{044}
&& V_{3}(\theta,\varphi) = [ {a} / {b} ]^{1 \over 2} { (\mathbb{I} - {a} -
6{b}) \over \bigl[ { \mathbb{I} - {a} - 3{b} } \bigr]^{1 \over 2}}
\gamma_{6} \cos(\theta - \varphi)\ + [ {a} / {b} ]^{1 \over 2} (\mathbb{I} -
{a} - 9{b}) \gamma_{8} \cos(\theta + \varphi)\ - \cr
&& {a} [{a} /{b}]^{1 \over 2} \gamma_{9} \cos(3\theta - \varphi)\ -\ {
3{a} \over [{\mathbb{I} - {a} - 3{b}} ] ^{1 \over 2}} \gamma_{7}
\cos(2\theta)
\end{eqnarray}
We choose \( a\ \And\ b \) in the domain \( \mathbb{B} \) defined by:
\begin{equation}
\mathbb{B} = \{(a, b) \in \mathbb{R}^{*2}_{+}\ ;\ a + 3b < \mathbb{I}\ ;\ \ |f|\ \&\
|V| < \varepsilon_{0} \}
\label{050}
\end{equation}
for the norm:
\begin{eqnarray}
\label{060}
&& |f| = \sup_{\nu \in \mathbb{Z}^{2}} |\hat{f}(\nu)|. F_{\sigma}(R.\nu) \\
&& \mbox{with}\ \ F_{\sigma}(\nu) = \sum_{j=0}^{\sigma} {|\nu|^{j} \over j!}
\label{059}
\end{eqnarray}
using the Fourier coefficients \( \hat{f} \) of \( f \). Likewise \(
|V| = |V_{1}| + |V_{3}| \).

The parameters \( \sigma,\ R,\ \varepsilon_{0} \) can be taken to be
\( \sigma = 3 \) (since we have 2 degrees of freedom), \( R = 1/3\ \
\And\ \ \varepsilon_{0} = 1/10 . \ \ \mathbb{B} \) is non-empty when \( \mathbb{I}
\leq 1/40 \). The shape of this domain is shown in the figure below.

The generators \( \alpha,\ G = (G_{1}, G_{3}),\ Z = (Z_{1}, Z_{3})
\) are given by (\ref{057}), (\ref{033}), (\ref{038}). And the
solution (\ref{031}), (\ref{055}) is written as:
\begin{eqnarray}
T^{-1}\left( \begin{array}{c} 0 \\ 0 \\ \theta \\ \varphi \end{array}
\right) = \left( \begin{array}{c} (\mathbf{1} + \dot{G})^{-1} \left(
\begin{array}{c} Z_{1} - \dot{\alpha}_{1} \\ Z_{3} - \dot{\alpha}_{3}
\end{array} \right) \\ \theta + G_{1} \\ \varphi + G_{3} \end{array}
\right)
\label{058}
\end{eqnarray}
with \( \ \dot{\alpha}_{1} = \partial_{\theta}\alpha\,,\
\dot{\alpha}_{3} = \partial_{\varphi}\alpha \). Let us recall that the
arguments of the functions \( \alpha_{1}, \alpha_{3}, G_{1}, G_{3} \)
are \( \theta, \varphi \). Then the flow acts on the angle variables
by (\ref{031}):
\begin{eqnarray}
\label{046}
&& \theta(t) = \theta(0) + \omega_{1} t \\
&& \varphi(t) = \varphi(0) + \omega_{3} t
\end{eqnarray}
since \( \hat{\omega} = \omega \) by the choice of \( Z \). To the
first order in \( \varepsilon \), we have (cf (\ref{057}),
(\ref{033})):
\begin{eqnarray}
\label{036}
&& \alpha = \Gamma f \\
&& G_{1} = \Gamma (V_{1}+ \Gamma \partial_{\theta}{f}) \label{063} \\
&& G_{3} = \Gamma (V_{3}+ \Gamma \partial_{\varphi}{f}) \label{064}
\end{eqnarray}
Now we define the Cantor set \( \mathbb{K} \) of Diophantine points \( (a,b)
\) by the conditions (\ref{016}) where \( \omega \) is actually an
affine function of \( a\ \And\ b\ \ (\mathbb{I} \) being fixed): cf
(\ref{061}), (\ref{062}). It is non-empty and has a positive Lebesgue
measure (if \( \gamma \) is small enough), as is well known. Let us
also define \( \widetilde{\mathbb{B}} = \mathbb{B} \cap \mathbb{K} \). We can also prove
that it is non-empty and has a positive Lebesgue measure (since \(
\mathbb{B} \) is an open set).

For any value of \( (a,b) \in \widetilde{\mathbb{B}} \) we can write a
trajectory of the Hamiltonian (\ref{014}) under the form
(\ref{046})-(\ref{064}) \& (\ref{012}). And so we get a trajectory of
the Hamiltonian (\ref{007}):
\begin{eqnarray}
\label{051}
&& \hat{a}_{1}(t) = \sqrt{a + Z_{1} - \Gamma \partial_{\theta}{f} +
O({\varepsilon}^{2}) }\ \exp i \Bigl( \theta + \Gamma V_{1}
+{\Gamma}^{2} \partial_{\theta}{f} + \xi + O({\varepsilon}^{2}) \Bigr)
\\
&& \hat{a}_{3}(t) =\sqrt{b + Z_{3} - \Gamma \partial_{\varphi}{f} +
O({\varepsilon}^{2})}\ \exp i \Bigl( \varphi + \Gamma V_{3}
+{\Gamma}^{2} \partial_{\varphi}{f}+ 3\xi + O({\varepsilon}^{2})
\Bigr) \\
&& \hat{a}_{2}(t) ={1 \over \sqrt{2}} \sqrt{\mathbb{I}- \hat{a}_{1}
\hat{a}_{1}^{*}-3\hat{a}_{3} \hat{a}_{3}^{*}}\ \exp (2i\xi)
\label{065}
\end{eqnarray}
with
\begin{eqnarray}
\partial_{t} \xi = \frac{\partial{H}}{\partial{\mathbb{I}}} \ \
(\mbox{ie}\ \ \xi = {\mathbb{I} \over 2}t + \xi_{0} + O({\varepsilon}^{2})\ )
\end{eqnarray}
\begin{figure}[hbt]
\epsfxsize=0.5 \hsize \epsfbox{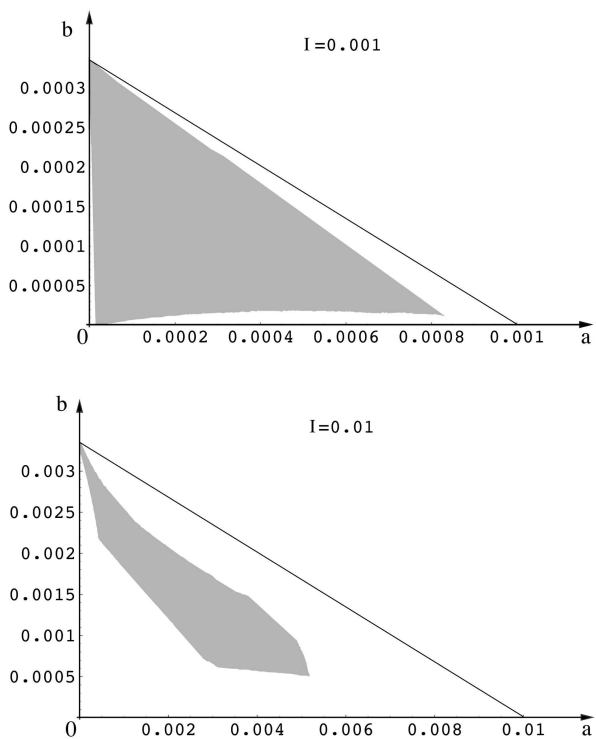} \caption{Domain \( \ \mathbb{B} \)
(shaded)\ when \( \ \mathbb{I} = 0.001\ \ \And\ \ \mathbb{I} = 0.01 \)}
\label{Fig1}
\end{figure}

We have seen in (\ref{052}) that \( \eta(x,t) \) is a travelling wave
if and only if:
\begin{equation}
\forall k,t \quad \ln(\hat{\eta}_{k}(t)) = e_{k} -i.k.c.t
\end{equation}
for some constants \( {e}_{k} \), which is equivalent, due to
(\ref{004}), to:
\begin{equation}
\label{037}
\forall k,t \quad \ln(\hat{a}_{k}(t)) = \hat{e}_{k} -i.k.c.t
\end{equation}
for some other constants \( \hat{e}_{k} \).

In particular taking \( k=1 \), in order to get a travelling wave, the
following should be true:
\begin{eqnarray}
\label{053}
\frac{1}{2} \ln(a + Z_{1}-\Gamma \partial_{\theta}{f}) + i(\theta +
\Gamma V_{1} +{\Gamma}^{2} \partial_{\theta}{f} + \xi) = \hat{e}_{1}-ict
\end{eqnarray}
But this cannot be true since the l.h.s formula is not affine in t
because \( f \) and \( V \) are quasi-periodic (cf
(\ref{042})-(\ref{044})). Hence our solutions are periodic in \( x \),
quasi-periodic in \( t \), and not travelling waves (since the
Diophantine vector \( \omega \) is not rational).

\section{Conclusion}
\label{S5}
\setcounter{equation}{0}

In this paper we considered the problem of existence of solutions of a
free surface in gravity waves, quasi-periodic in time and periodic in
space. We applied the KAM method to the particular Zufiria model of 3
waves interaction which is a truncation of the Zakharov Hamiltonian.
From the expressions (\ref{051})-(\ref{065}) we remark that these
solutions are not the ones obtained by the Hopf bifurcation, as in
(Van der Meer, 1985). For the Hopf bifurcation, the solution is
obtained by a continuous variation of the parameter, which is the
celerity, of a permanent form solution. This is not the case for the
solutions given by the KAM method, they exist in a domain of the phase
space which contains the non-empty Cantor set \( \widetilde{\mathbb{B}} \).
Moreover, these solutions are not null solutions since \( \alpha\ \&\
G \) are not identically zero. And they are not permanent form
solutions neither because \( \alpha \) and \( G \) are not constant
functions.

This method may be extended to a finite trucation of the Zakharov
Hamiltonian with more modes involved.

In extension to this work, we would like to give a better threshold
for the domain of the existence of the solution. Our next aim,
unsolved for the moment, would be the case of a system of infinitely
many degrees of freedom. It is known to be a difficult problem in KAM
theory.

\end{document}